\documentclass[PRB,twocolumn,showpacs,floatfix]{revtex4-1}
\usepackage{graphicx,epsfig,psfrag}
\usepackage{epstopdf}
\usepackage{amsfonts}
\usepackage{times}
\usepackage{graphics,dcolumn,bm,epic,eepic,float}
\usepackage{amssymb,amsmath,amsfonts,rotate,color}
\usepackage[colorlinks=true,linkcolor=blue,urlcolor=blue,citecolor=blue,anchorcolor=blue]{hyperref}
\setcitestyle{numbers,square}

\begin{document}
\newcommand{\be}{\begin{equation}}
\newcommand{\ee}{\end{equation}}
\newcommand{\bearr}{\begin{eqnarray}}
\newcommand{\eearr}{\end{eqnarray}}
\newcommand{\nn}{\nonumber}
\newcommand{\la}{\langle}
\newcommand{\ra}{\rangle}
\newcommand{\cd}{c^\dagger}
\newcommand{\vd}{v^\dagger}
\newcommand{\fd}{f^\dagger}
\newcommand{\tk}{{\tilde{k}}}
\newcommand{\tp}{{\tilde{p}}}
\newcommand{\tq}{{\tilde{q}}}
\newcommand{\eps}{\varepsilon}
\newcommand{\vk}{{\vec k}}
\newcommand{\vp}{{\vec p}}
\newcommand{\vq}{{\vec q}}
\newcommand{\vkp}{\vec {k'}}
\newcommand{\vpp}{\vec {p'}}
\newcommand{\vqp}{\vec {q'}}
\newcommand{\bk}{{\bf k}}
\newcommand{\bp}{{\bf p}}
\newcommand{\bq}{{\bf q}}
\newcommand{\up}{\uparrow}
\newcommand{\down}{\downarrow}
\newcommand{\fns}{\footnotesize}
\newcommand{\cdag}{c^{\dagger}}
\newcommand{\lc}{\langle\!\langle}
\newcommand{\rc}{\rangle\!\rangle}

\title{Majorana bound states in magnetic impurity chains: Effects of $d$-wave pairing}
\date{\today}
\author{Mahdi Mashkoori}
\author{Annica Black-Schaffer}
\affiliation{Department of Physics and Astronomy, Uppsala University, Box 516, SE-751 20 Uppsala, Sweden}

\begin{abstract}
We consider an atomic chain of magnetic impurities on the surface of a spin-orbit coupled superconductor with a dominating $d$-wave and subdominating $s$-wave order parameters. In particular, we investigate the properties of the Majorana bound states (MBSs) emerging at the chain end points in the topological phase and how MBSs are affected by the $d$-wave order parameter. We provide a comprehensive picture by both studying time-reversal invariant and breaking superconducting substrates as well as chains oriented in different directions relative to the $d$-wave rotation.
We show that increasing the $d$-wave order parameter significantly enhances the localization of MBSs and their protective minigap, as long as the direction along which the impurity chain is oriented does not cross any nodal lines of the gap function. 
Moreover, we find an extra gap-closing for a specific condensate and chain orientation within the topological phase, which we are able to attribute to simple geometrical effects in the corresponding two-dimensional limit. 
These results show how high-temperature $d$-wave superconductors can be used to significantly enhance the properties and stability of MBSs.
\end{abstract}

\maketitle
%
%
\section{Introduction}
Topological superconductivity generating Majorana bound states (MBSs) in low dimensional systems represent one of the most spectacular quantum states in condensed matter physics \cite{Kitaev2001,Fu-Kane2008,Wilczek2009,Sato2009,QiRMP2011,Alicea2012,Leijnse2012,BeenakkerRMP2015}. During the last few years, several platforms for engineering topological superconductors (SCs) and detecting MBSs have been developed. Among them, magnetic impurities on top of a spin-orbit coupled SC has shown great promise and versatility \cite{Mourik2012, Anindya2012, Churchil2013, Pawlak2015,Yazdani2014, Ruby2017, Menard2017, Wiesendanger2018,Kamlapure18}. In particular, one-dimensional (1D) atomic chains of magnetic impurities on the surface of conventional $s$-wave SCs with Rashba spin-orbit coupling have been studied intensively \cite{Choy2011,Klinovaja2013,Nadj-Perge2013, Oppen2014, Pientka2013, DasSarma2012, DanielLoss2012}. These MBSs are robust against non-magnetic disorder \cite{Awoga2017} and their emergence is also not restricted to single impurity chains, but MBSs also appear at odd-numbered junctions in impurity chain networks \cite{KristoferRapidC2016}. Extending also to two dimensions (2D), Majorana edge states has been investigated around whole islands of magnetic impurities \cite{Menard2017,Kristofer2018}. Throughout all these studies, the superconducting substrate has been a conventional $s$-wave SC.

The unconventional $d$-wave cuprate SCs offer a tantalizing possibility to realize MBSs at much higher temperatures thanks to their larger order parameter \cite{Takei2013,Ortiz2018,wang2013} and higher transition temperatures. However, for $d$-wave SCs the absence of a full energy gap appears to pose an insurmountable obstacle as nodal quasiparticles pollute the low-energy spectrum, hybridize with the MBSs, and thus destroy their protection. Also, in terms of magnetic impurities, $d$-wave SCs only host resonance states with a finite life-time, i.e.,~virtual bound states \cite{Pan2000,Hudson2001,Zhou2013,Balatsky1995,BalatskyRMP2006}. This is in sharp contrast with the magnetic impurity induced subgap bound states in $s$-wave SCs, the so-called Yu-Shiba-Rusinov (YSR) states \cite{Yu1965,Shiba1968,Rusinov1969}, which are the building blocks of topological SCs in magnetic impurity-based platforms.

The problem with nodal quasiparticles in $d$-wave SCs could potentially be resolved if a coexisting but subdominant $s$-wave order parameter is also present. There exists some evidence for such coexistence of dominant $d$-wave and subdominant $s$-wave order parameter in cuprate SCs \cite{KirtleyRMP, Kirtley2005, Schemm2014,Hakansson2015, Schemm2015, Gonge2017}. 
For example, fully gapped $d$-wave SCs has been found at specific surfaces and also in nanoislands of YBa$_2$Cu$_3$O$_{7-\delta}$ \cite{Lombardi2013,Black-Schaffer2013}. More generally, in the vicinity of interfaces a time-reversal symmetry breaking order parameter is anticipated to appear in $d$-wave SCs \cite{Matsumoto95I,Matsumoto95II,Fogelstrom97,Sigrist98}. In addition,  
an engineered alternative is a hybrid structure between an unconventional $d$-wave SC and an atomically thin layer of conventional $s$-wave SC, which can produce a superconducting state combining the benefit of the high transition temperature of the $d$-wave SC with an additional $s$-wave component. 
In general, the  order parameter in these systems takes the form $\Delta = \Delta_d + e^{i\alpha} \Delta_s$, where $\alpha =0$ gives a time-reversal invariant (TRI) phase, while $\alpha =\pi/2$ results in a time-reversal broken (TRB) phase.
In the TRB $d+is$-wave SC, the coexistence gives rise to a fully gapped spectrum where a single magnetic impurity then induces YSR-subgap states \cite{Mahdi2017}. On the other hand, a TRI $d+s$ SC with a dominating $d$-wave order still has nodal lines, although modified from the pure $d$-wave state.

In this paper, we investigate if and how a coexisting $s$-wave order can turn high-temperature $d$-wave SCs into a viable platform for MBSs forming at the end of magnetic impurity chains. We assume dominating $d$-wave order, consider both TRI and TRB coexistence phases with a small $s$-wave component, and study chains oriented in different directions on the substrate relative to the $d$-wave rotation, all to provide a comprehensive study.

First, we show that MBSs actually emerge for an impurity chain embedded in TRI SC with $d{+}s$-wave symmetry, despite the nodal lines in the order parameter. However, it requires some tuning, especially of the doping level. Also, there is a strong dependence on chain orientation relative to the $d$-wave rotation: If the impurity chain crosses the nodal lines of the order parameter, the minigap which protects the MBSs from quasiparticle excitations is strongly suppressed. Besides the emergence of MBSs and the protecting minigap, we also focus on the localization of the MBSs. We show that the localization length of the MBSs depends on effective order parameter along the chain, not necessarily the minigap. For all viable chain orientations, we find that the $d$-wave component significantly enhances the MBS localization and the minigap compared to a pure $s$-wave substrate. 

Next we study an impurity chain on a TRB $d{+}is$-wave SC. The complex order parameter results in a full energy gap in the excitation spectrum, which results in the appearance of MBSs becoming largely parameter independent, as well as directional independent. Importantly, the $d$-wave component strongly enhances the minigap and MBS localization. The only exception is the TRB $d_{xy}{+}is$-wave SC, where we find an extra gap-closing for $y$-axis chains, but not for $x$-axis chains. We show that the extra gap-closing is not due to any additional topological phases or phase transitions, but are the result of flat chiral edge states in the 2D limit. This demonstrates that sample geometry can, in fact, overshadow topology in determining the boundary spectrum.
It is here worth mentioning that spin-orbit coupled nanowires on top of a pure $d$-wave SC have also recently been studied. However,  the emergence of a topologically non-trival superconducting phase was in that setup predicted to be strongly dependent on the relative orientation of the nanowire and $d$-wave order \cite{Takei2013} and the MBS localization to be reduced compared to a $s$-wave substrate \cite{Ortiz2018}. By using a combination of $d$- and $s$-wave orders, we are able to circumvent both of these issues.
In summary, our results for magnetic impurity chains demonstrates that a high-temperature $d$-wave SC can dramatically enhance the properties of MBSs, including both significantly increased minigaps and shorter MBS localization lengths, as soon as a small coexisting $s$-wave state is present. Notably, this result does not depend on the relative phase between the $d$- and $s$-wave components, making our results generally applicable, independent of details of the superconducting state. 
 
The reminder of this paper is organized as follows. In Sec.~\ref{section_model}, we introduce the numerical tight-binding lattice model used to study $d$- and $s$-wave substrates with magnetic impurity chains. In Sec.~\ref{section_results}, we present our results where we focus on how $d$-wave pairing affects the MBSs at the impurity chain end points. We explain both how different chain orientations and different $d$-wave order parameter rotations influence the results. We present complementary discussions and a short comparison to the pure $d$-wave and nanowire platform in Sec.~\ref{section_discussion} and, finally, in Sec.~\ref{section_conclusion} we summarize our results.

%
\section{Model}{\label{section_model}}
\subsection{Superconducting substrate}
To model impurity chains with emergent MBSs, we consider a system with a spin-orbit coupled superconducting substrate. To easily incorporate different superconducting pairing symmetries while still keeping the model as simple as possible, we  consider a mean-field Bogoliubov-de Gennes (BdG) Hamiltonian given by
\begin{align}
& H_{\rm sub} =  -t \sum\limits_{\left\langle {{\bf{i}},{\bf{j}}} \right\rangle \sigma } {c_{{\bf{i}}\sigma }^\dag {c_{{\bf{j}}\sigma }}} -\mu \sum\limits_{\bf{i} \sigma} {c_{{\bf{i}}\sigma }^\dag {c_{{\bf{i}}\sigma }}} \nonumber\\ 
& -\lambda_R \sum\limits_{\bf i ,\eta=\pm} \eta c_{{\bf i}\uparrow}^{\dag} 
(c_{{\bf i}-\eta\hat{\bf x}\downarrow}-
ic_{{\bf i}-\eta\hat{\bf y}\downarrow})+ \rm{ H.c.}  \nonumber \\ 
& + \! \sum\limits_{\bf{ij}} {[{\Delta _s} \left( {\bf{i}} \right)\delta_{\bf{ij}}+\frac{1}{4} {\Delta _d\left( {{\bf{i}},{\bf{j}}} \right)}]}c_{{\bf{i}} \uparrow }^\dag c_{{\bf{j}} \downarrow }^\dag + \rm{ H.c.},
\label{BdG1}
\end{align}
where $c_{\bf i} (c^\dagger_{\bf i})$ is creation (annihilation) operator at $\textbf{i} = (i_x,i_y)$, which represents a site in a square lattice, with the lattice spacing $a$ set to be $1$.  Here $\mu$ represents the chemical potential, while $t$ is the hopping matrix element to the nearest neighbors. We also add Rashba spin-orbit interaction in the substrate set by $\lambda_R$, which is always present due to inversion symmetry breaking at the SC surface.
For superconductivity, we assume both $d$-wave and $s$-wave pairing. The $d_{x^2-y^2}$-wave ($d_{xy}$-wave) order can be modeled to exist on nearest- (next-nearest) neighboring bonds, while the conventional $s$-wave order is an on-site parameter.  In most calculations, we keep the order parameter constant, i.e.,~non-self-consistent calculations, where we enforce $d_{x^2-y^2}$-wave order by setting  $\Delta_d((i_x,i_y),(i_x\pm1,i_y)) = -\Delta_d((i_x,i_y),(i_x,i_y\pm1))$ for all sites. For the $d_{xy}$-wave order, we follow the same procedure but on the diagonal bonds instead.
The coexistence of $d$-wave and $s$-wave order parameters has been observed in several materials and for generality we consider $\Delta = \Delta_d + e^{i\alpha} \Delta_s$, where $\alpha$ is set to be either $\alpha = 0$ or $\pi/2$. This captures both all fully real condensates ($\alpha =0$) and TRB cases ($\alpha = \pi/2$). The latter is generally favored if external factors do not prevent a full relaxation of the superconducting order since it has a fully gapped spectrum.

\subsection{Gap function nodal lines and Fermi surfaces}
\begin{figure}[t]
\center
\includegraphics[width=0.27\textwidth]{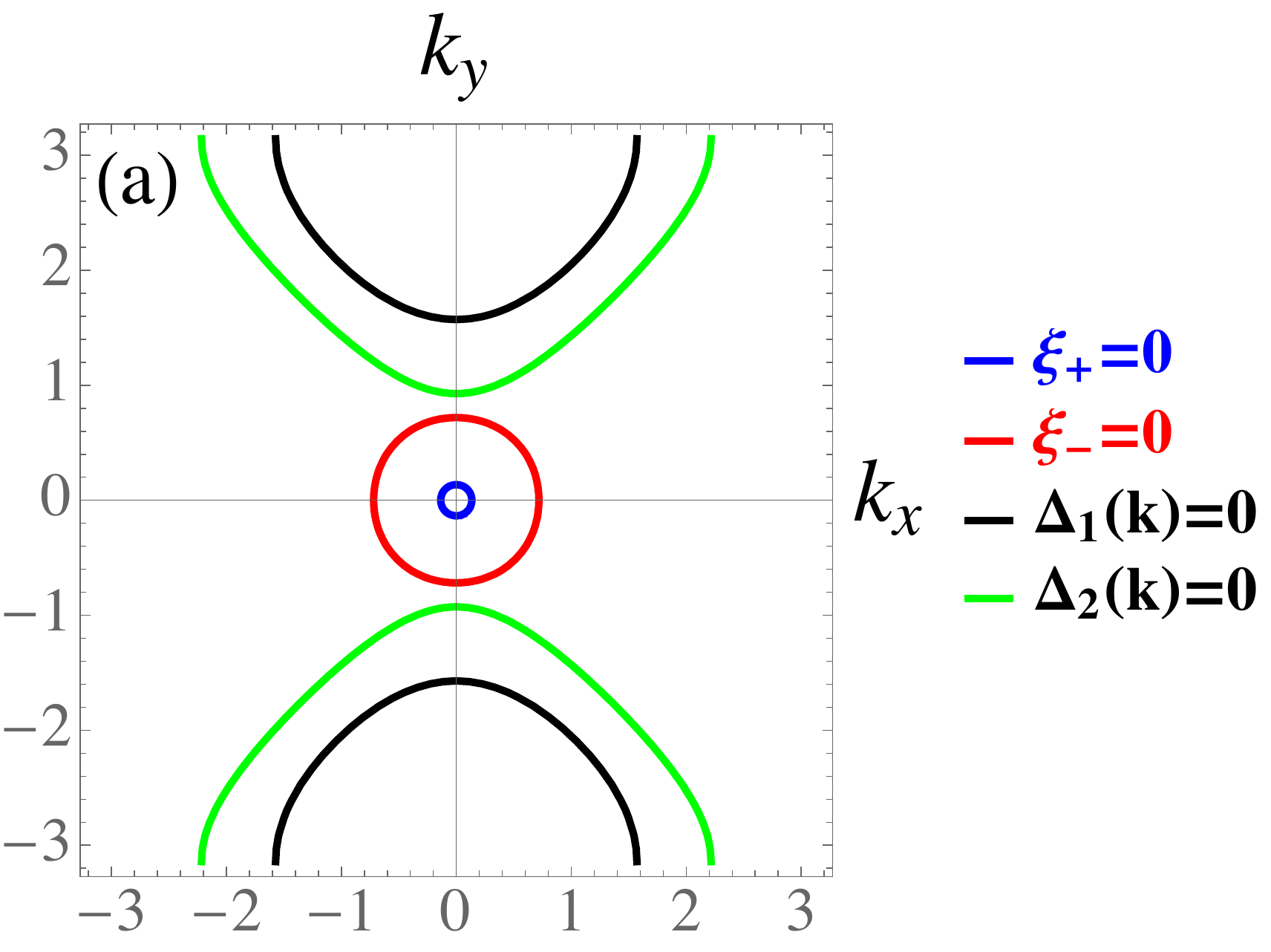}
\includegraphics[width=0.2\textwidth]{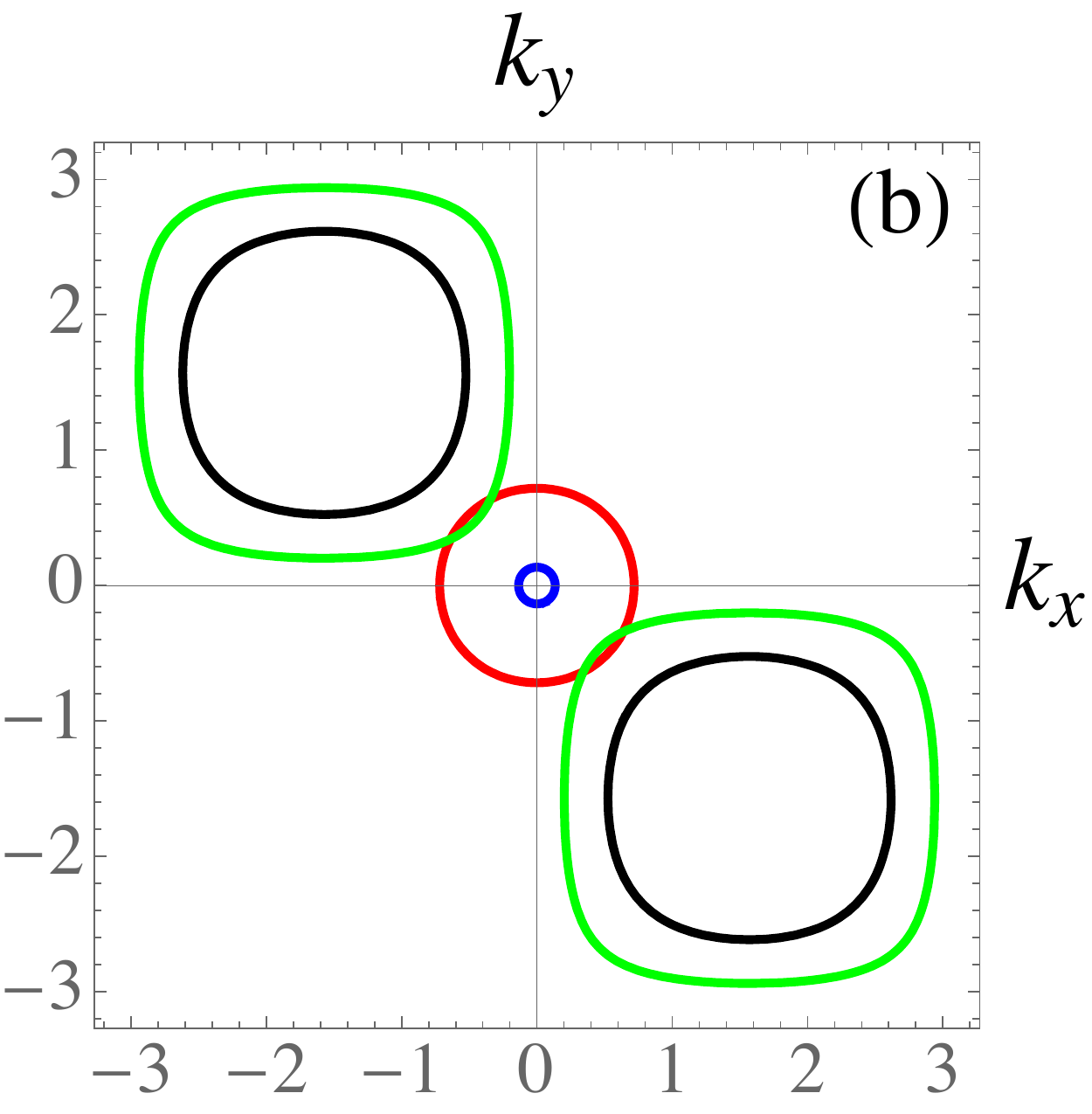}
\caption{(Color Online) Nodal lines of the order parameter, $\Delta(k)=0$, (black, green) for $d_{x^2-y^2}{+}s$-wave (a) and $d_{xy}{+}s$-wave states (b) with Fermi surfaces of the normal Hamiltonian, $\xi_\pm(k)= 0$ (blue, red). Here $\Delta_s = 0.1$ and $\mu = -3.9$. For $\Delta_1(k)$ (black) and $\Delta_2(k)$ (green), we set $\Delta_d$ to be $0.2$ and $0.5$, respectively. } 
\label{NodalLine.fig}
\end{figure}

All TRB superconducting substrates, $d_{x^2-y^2}{+}is$- and $d_{xy}{+}is$-wave SCs, have a full energy gap in the spectrum and thus the order parameter does not have any nodal lines. However, for the TRI solutions, the gap structure depends in more detail on the parameters.
We start with exploring the TRI $d_{x^2-y^2}{+}s$-wave superconducting substrate. After performing a Fourier transform, the superconducting order parameter in reciprocal space reads $ \Delta(k) = \Delta_d(\cos k_y-\cos k_x)/2+\Delta_s$, where we assume both $\Delta_s$ and $\Delta_d$ are positive definite without loss of generality. This order parameter contains an isotropic $s$-wave and a sign-changing anisotropic $d$-wave order parameters. As long as the $s$-wave component is the dominant order, i.e.,~$\Delta_s > \Delta_d$, the gap function does not have any nodes and the spectrum must be fully gapped. 
However, as shown in Fig.~\ref{NodalLine.fig}(a), when the $d$-wave order is dominating, $\Delta_s < \Delta_d$, the gap function in the first Brillouin zone changes sign and nodal lines appear in the gap function ($\Delta(k) =0$ curves in black and green). In fact, the figure illustrates the anisotropy of the order parameter, which will explain the MBS's dependence on the impurity chain orientation as found in the numerical results. 
Next, we focus on the $d_{xy}{+}s$-wave SC and, in this case, where the Fourier-transformed order parameter reads $\Delta(k) = \Delta_{d}(\sin k_x \sin k_y)+\Delta_s$. Similar analysis as above gives the modified nodal lines of the order parameter as depicted in Fig.~\ref{NodalLine.fig}(b) when the $d_{xy}$-wave order is largest. 

In Fig.~\ref{NodalLine.fig}, we also depict the Fermi surfaces of the normal Hamiltonian (blue and red curves) for lightly electron doped bands at $\mu = -3.9$. Here, the Rashba spin-orbit interaction splits up the spin-degenerate bands into two helical bands, $\xi_{\pm}(k) = -2t(\cos k_x + \cos k_y)-\mu \pm 2\lambda_R \sqrt{\sin^2 k_x +\sin^2 k_y}$. 
Being interested in dominant $d$-wave superconductivity but still requiring a fully gapped spectrum, a prerequisite for YSR-states to emerge, we need the Fermi surfaces to not cross the nodal lines. As a consequence, for all TRI cases, we have to consider the Fermi level to be at the bottom or top of the band $|\mu| \approx 4$ and  avoid too large $\Delta_d/\Delta_s$ ratios. As an example, for $\Delta_d/\Delta_s=5$ (green) in Fig.~\ref{NodalLine.fig}, the nodal lines almost touch the Fermi surface and the excitation spectrum becomes gapless, even at light doping. Actually, even with these limitations, the spectrum fails to be gapped if the Rashba spin-orbit interaction $\lambda_R$ becomes the largest energy scale, as that separates the spin degenerate Fermi surfaces and thus pushes the outer Fermi surface toward order-parameter nodal lines. Still, based on physically relevant parameter regimes,  there exist some regions in the parameter space for which the spectrum is fully gapped for the TRI solutions and we set spin-orbit coupling and chemical potential in a way to comply with these restrictions.

\subsection{Impurity chain}
To model an impurity chain, we assume the spin of each impurity to be a classical vector which effectively acts as a local Zeeman field \cite{Nadj-Perge2013,Pientka2013}:
\be
 H_{\rm imp} = \sum\limits_{\textbf{R}\sigma \sigma'}  J \vec{S_{\textbf{R}}} \cdot  c_{\textbf{R} \sigma}^{\dag} (\vec{\sigma})_{\sigma \sigma'} c_{\textbf{R} \sigma'},
\ee
where $\vec{S}_R$ is the impurity spin and $J$ represents the exchange coupling between each impurity and the superconducting substrate.
The impurity chain can be spatially oriented either along the $x$- or $y$-axis and is always placed in the middle of the square lattice to avid any possible influence from the boundaries.
Working in the classical limit, we set $|\vec{S}| \to \infty$ while $J \to 0$ in the manner that  $U_{\rm mag} \equiv J|\vec{S}|$ remains finite. 
To drive the chain into the topological phase, we need to assume either a ferromagnetic impurity chain with Rashba spin-orbit interaction in the substrate or a spin helical impurity chain. In this work, we mainly set the impurity chain to be ferromagnetic and we include Rashba spin-orbit interaction in the substrate. The only exception is Sec.~\ref{subsection_spin-helix} where we exclude the substrate spin-orbit interaction and instead assume a helical structure for the local moments of impurities.
Since the hopping $t$ in $H_{\rm sub}$ is also active between impurity chains, it can be seen as either the hopping in the substrate or between the impurities. In this way we capture within a single simple model qualitatively both the Shiba band and the ferromagnetic wire limits \cite{Hoffman2016, Andolina2017, Theiler2018}.

In this paper, all the energies are scaled by the nearest neighbor hopping matrix element; $t=1$. Moreover, for the ferromagnetic impurity chain we assume that the spins are along $z$ direction and we fix the spin-orbit interaction to be $\lambda_R = 0.3$.
The Rashba spin-orbit interaction for an Fe chain on top of Pb has been estimated to $\lambda_R = 0.05$ eV, while the hopping to the nearest neighbors for different orbitals of the Fe atom ranges from $0.1$ eV to $0.7$ eV\cite{Yazdani2014}. Thus, the assumption of $\lambda_R/t = 0.3$ is realistic. We have also verified that our general conclusions are insensitive to the exact parameter values.
In all non-self consistent calculations the on-site $s$-wave order parameter is set to $\Delta_s=0.1$, while $d$-wave order is tuned; $0.1~<~\Delta_d~<~2$, to allow to study the impact of varying but dominating $d$-wave orders. We obtain the eigenvectors and eigenvalues based on diagonalization of Hamiltonian $H = H_{\rm sub} + H_{\rm imp}$ in real space within the BdG framework. For the superconducting substrate we choose a square lattice with dimensions $L_{\parallel} \times L_{\perp}$ lattice points where along the impurity chain $501 \le L_\parallel \le 1001$ and perpendicular to it $11 \le L_\perp \le 51$. The impurity chain is $\frac{3}{5} L_\parallel$-sites long and laying in the middle of the substrate. In diagonalizing the tight-binding Hamiltonian, we utilize the Arnoldi iteration
 scheme from TBTK toolkit \cite{TBTKtoolkit,Bjornson18}. 

%
%
\section{Results}{\label{section_results}}
Having defined a general model to study the influence of $d$-wave pairing in the previous section, we here report the results. 
In Secs.~\ref{subsection_d1+s} and {\ref{subsection_d2+s}, we study a ferromagnetic impurity chain on a general TRI SC substrate and describe how the properties of MBSs, such as localization length scale and minigap energy, are affected by the additional $d$-wave order parameter and compare it with pure $s$-wave case. Considering the symmetry of $d$-wave component of the order parameter, we discuss $d_{x^2-y^2}$-wave and $d_{xy}$-wave state in two different subsections. 

In Secs.~\ref{subsection_d1+is} and \ref{subsection_d2+is}, we instead consider the opposite type of coexistence of $d$- and $s$-wave order, by allowing the SC to break time-reversal symmetry. 
Finally, in Sec.~\ref{subsection_self-consistency}, we discuss the effect of self-consistent calculations for the order parameters and also consider spin helical impurity chains in Sec.~\ref{subsection_spin-helix}.  

\subsection{$d_{x^2-y^2}{+}s$-wave substrate}{\label{subsection_d1+s}}
To understand the overall behavior of a TRI $d_{x^2-y^2}{+}s$-wave substrate, we use Fig.~\ref{NodalLine.fig}(a) and perform dimensional reduction by temporarily setting $k_y=0 (k_x=0)$ for an impurity chain oriented along the $x$-axis ($y$-axis) \cite{Mynote1,Kristofer2015}. Consequently, for an $x$-axis impurity chain on a $d_{x^2-y^2}{+}s$-wave SC, the order parameter reads $\Delta(k_x) = \Delta_d(1-\cos k_x)/2+\Delta_s$ and, clearly, $\Delta_s \le \Delta(k_x) \le 2\Delta_d+\Delta_s$, which means that the $d$-wave order enhances the total order parameter along the impurity chain. On the other hand, for a $y$-axis impurity chain, the total order parameter reads $\Delta(k_y) = \Delta_d(\cos k_y-1)/2+\Delta_s$, which leads to $-2\Delta_d+\Delta_s \le \Delta(k_y) \le \Delta_s$. Obviously, in this latter case the total order parameter has nodal points, $\Delta(k_y)=0$, where the superconducting order parameter will be strongly suppressed anywhere near these nodes. Therefore, we anticipate very different behavior for $x$- and $y$-axis impurity chains in $d_{x^2-y^2}{+}s$-wave SCs. 

\begin{figure}[t]
\centering
\includegraphics[width=0.5\textwidth]{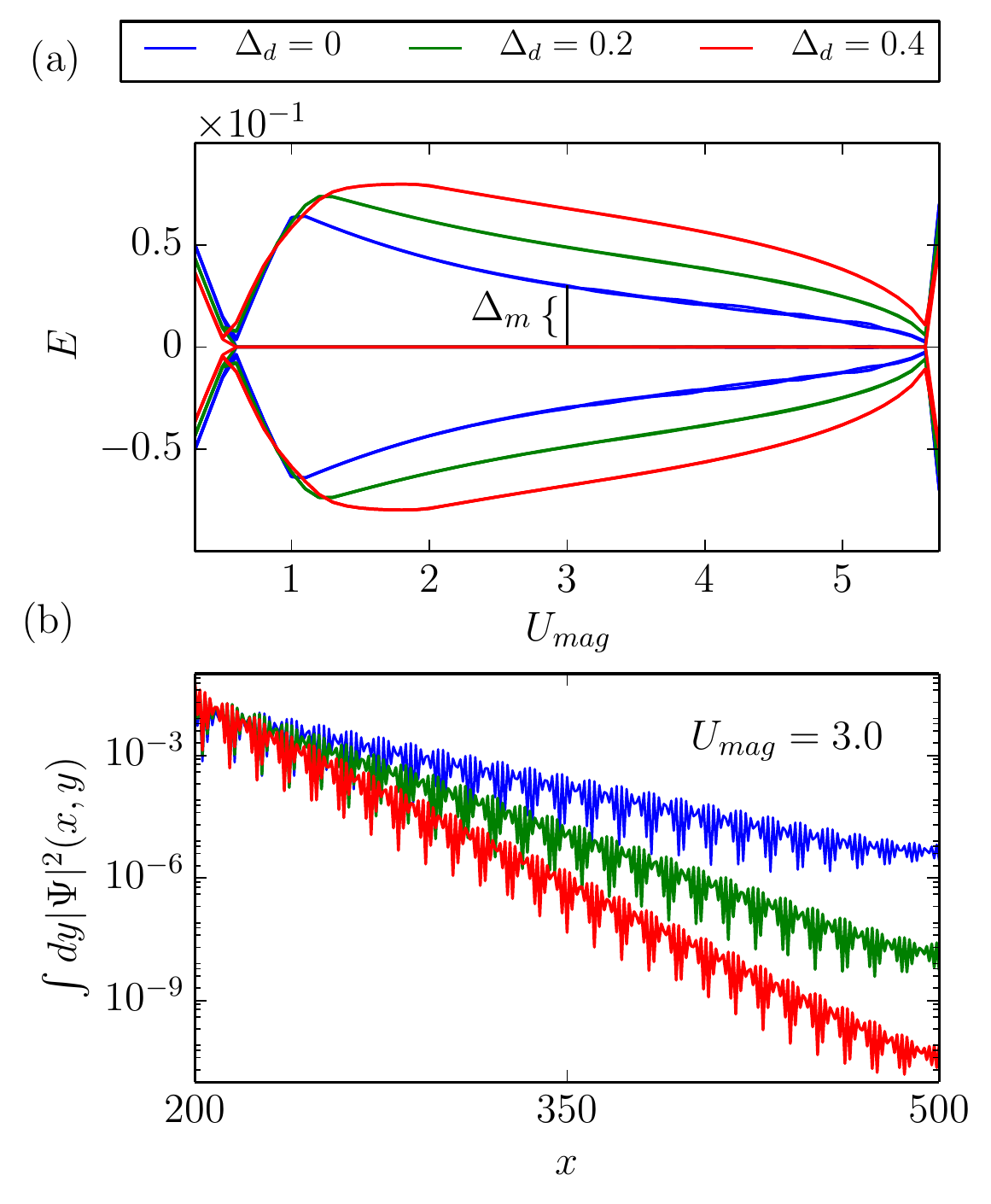}
\caption{(Color Online) Energy of lowest subgap states for $d_{x^2-y^2}{+}s$-wave SC with impurity chain oriented along $x$-axis (a) and magnitude squared wave function of MBSs along impurity chain in logarithmic scale (b). Blue curve represents a reference with only $s$-wave order. Here $\lambda_R = 0.3$ and $\mu = -4.0$.}
\label{d1+s_x.fig}
\end{figure}
In Fig.~\ref{d1+s_x.fig}(a), we plot the energy of the lowest energy subgap states for an $x$-axis impurity chain embedded in a 2D $d_{x^2-y^2}{+}s$-wave SC as a function of $U_{\rm mag}$, the strength of magnetic interaction between the impurities and SC. Given that each magnetic impurity induces a pair of YSR-subgap states, for the impurity chain many subgap states emerge. By increasing $U_{\rm mag}$ from very small values to a critical $U^{(1)}_{mag}$, these YSR states move deeper inside the gap and eventually touch each other at the Fermi level. This gap closing is the topological phase transition in the system and a pair of MBSs emerges at the impurity chain end points. In this particular, case we see that the MBSs emerge for $U^{(1)}_{mag} \approx 0.5$ and disappear for $U^{(2)}_{mag} \approx 5.8$, thus, the system is in topological nontrivial phase in between. Thermal hybridization of the MBSs with other states is protected by the minigap $\Delta_{m}$, the energy barrier between MBSs and the first excited state \cite{Sau2010,Alicea2012}. Having the impurity chain oriented along the $x$-axis, the coexistence of $d$-wave and $s$-wave order parameters turns out to be highly beneficial. As seen in Fig.~\ref{d1+s_x.fig}(a), the minigap $\Delta_{m}$ increases with increasing  $d_{x^2-y^2}$-wave component (green and red), compared to pure $s$-wave (blue). We also see that the critical couplings $U^{(1)}_{mag}$ and $U^{(2)}_{mag}$ for the topological phase transitions do not show any significant change when increasing $\Delta_d$. 

\begin{figure}[t]
\includegraphics[width=0.5\textwidth]{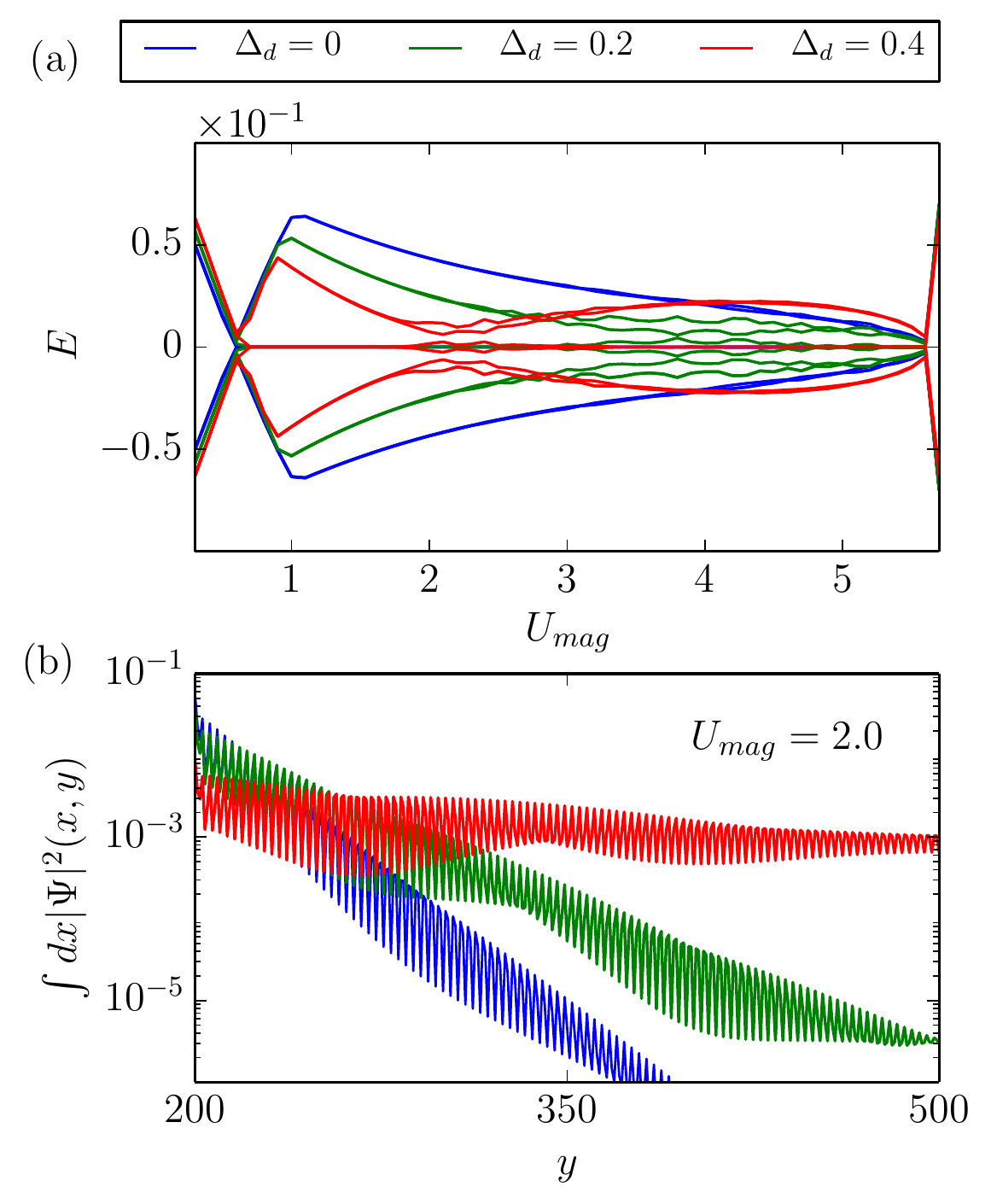}
\caption{(Color Online) Same as Fig.~\ref{d1+s_x.fig} but for impurity chain oriented along $y$-axis.}
\label{d1+s_y.fig}
\end{figure}
Another important impact of the $d$-wave order parameter on the MBSs is an increase in the MBSs localization as depicted in Fig.~\ref{d1+s_x.fig}(b). In this figure, we plot the magnitude squared wave function of the lowest energy states, the MBS, as a function of the $x$ coordinate along the chain from the chain end point in a logarithmic scale. The exponential decay of MBSs is obvious, with an additional oscillatory envelop related to Fermi wave vector $k_F$ \cite{DasSarma2012,DanielLoss2012,Pientka2013}. Thus the localization of MBSs is strongly enhanced due to the presence of $d$-wave order parameter compared to pure $s$-wave SC. The reason behind this localization enhancement is that the nodal lines of the $d_{x^2-y^2}{+}s$-wave state do not to cross the $k_x$ axis in the first Brillouin, zone which results in an overall larger energy gap along $k_x$ and thus more isolated MBSs. However, if we increase $\Delta_d$ a lot more beyond $\Delta_d/\Delta_s \sim 5$, the energy gap starts to shrink as the order parameter nodal lines eventually approach the Fermi surfaces and, consequently, the size of minigap is reduced. 
Therefore, to have reasonably robust and localized MBSs, there is an upper limit to the enhancement produced by a $d$-wave order in the $d_{x^2-y^2}{+}s$-wave substrate.

Next we consider a $y$-axis impurity chain for the same $d_{x^2-y^2}{+}s$-wave substrate. As shown in Fig.~\ref{d1+s_y.fig}(a), the coexistence of $d$-wave and $s$-wave pairing now leads to a much smaller minigap in major regions of the topological phase. As for the localization of MBSs, we show in Fig.~\ref{d1+s_y.fig}(b) that the localization is also strongly suppressed with increasing $d$-wave order parameter.  Following the same way of reasoning as for the impurity chain along the $x$-axis, the suppression of minigap for the $y$-axis impurity chain is easily attributed to the nodal gap of the order parameter along the chain orientation, as seen in Fig.~\ref{NodalLine.fig}(a). 
We find, thus, that the coexistence of $d_{x^2-y^2}$-wave and $s$-wave pairing suppresses or enhances the minigap and localization length scale of MBSs, depending on the impurity chain orientation with respect to anisotropic $d$-wave order parameter.

\begin{figure}[tb]
\includegraphics[width=0.45\textwidth]{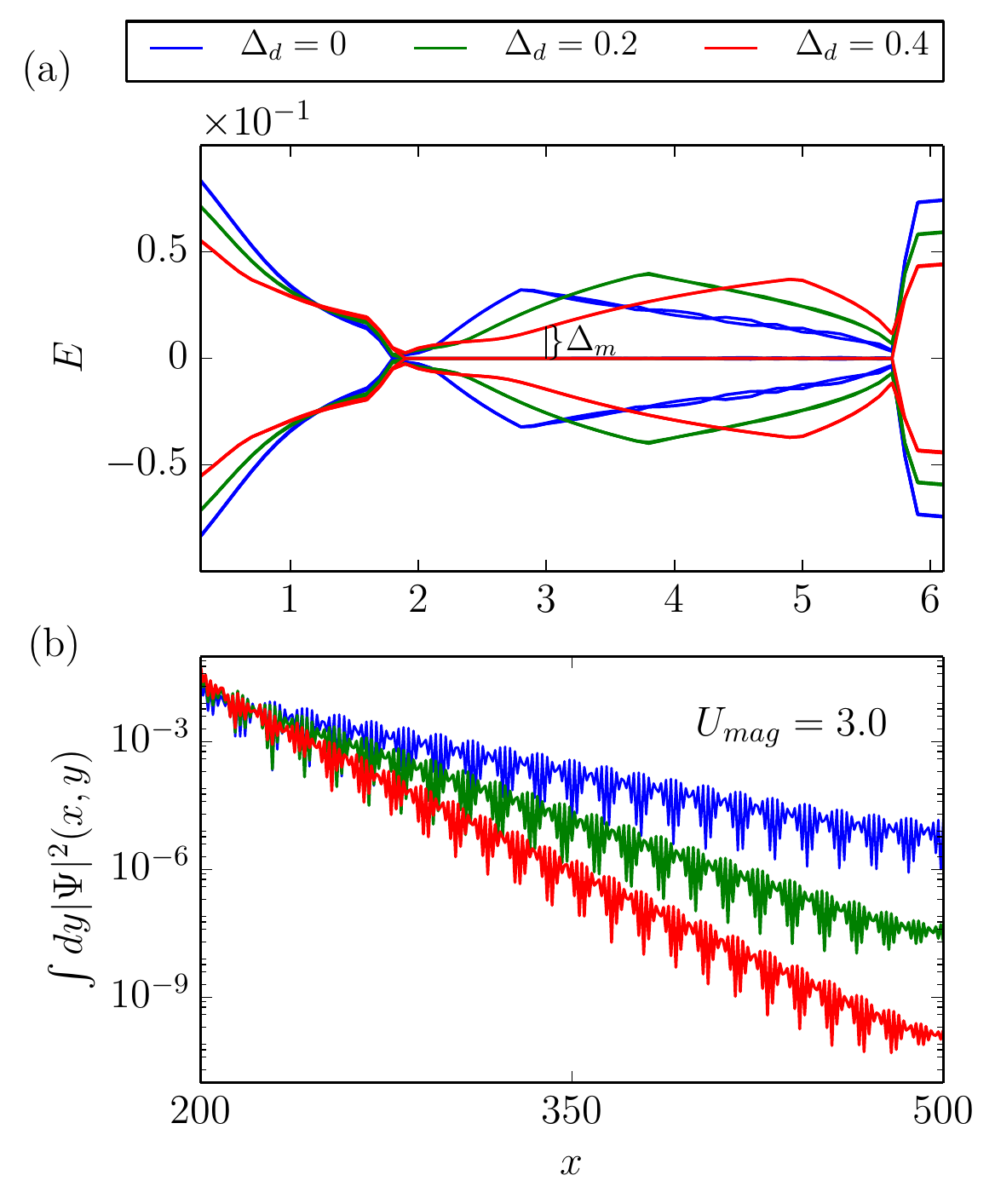}
\caption{(Color Online) Same as Fig.~\ref{d1+s_x.fig} but when also including next-nearest-neighbor hopping $t' =0.1$.}
\label{D+S_nnnHopping.fig}
\end{figure}
One might tend to naively directly relate the enhancement (suppression) of MBSs localization to having larger (smaller) minigap. However, we find that this is not always true, as illustrated in Fig.~\ref{D+S_nnnHopping.fig}. Taking into account next-nearest-neighbor hopping $t'=0.1$ for a $d_{x^2-y^2}{+}s$-wave SC substrate, we show in Fig.~\ref{D+S_nnnHopping.fig}(a) that for a $x$-axis impurity chain, increasing the $d$-wave order parameter can also lead to a smaller minigap than that for a pure $s$-wave SC. Still in the same parameter regime, the localization of MBSs is enhanced with increasing $d$-wave order. In fact, the localization of MBSs for finite $t'$ as depicted in Fig.~\ref{D+S_nnnHopping.fig}(b) gives a very similar result to the $t'~=~0$ case as seen in Fig.~\ref{d1+s_x.fig}(b), while the minigap energies are very different. Therefore, we find that the localization length of MBSs is more determined by the superconducting order parameter itself and the impurity chain orientation rather than the minigap.
\begin{figure}[t]
\includegraphics[width=0.45\textwidth]{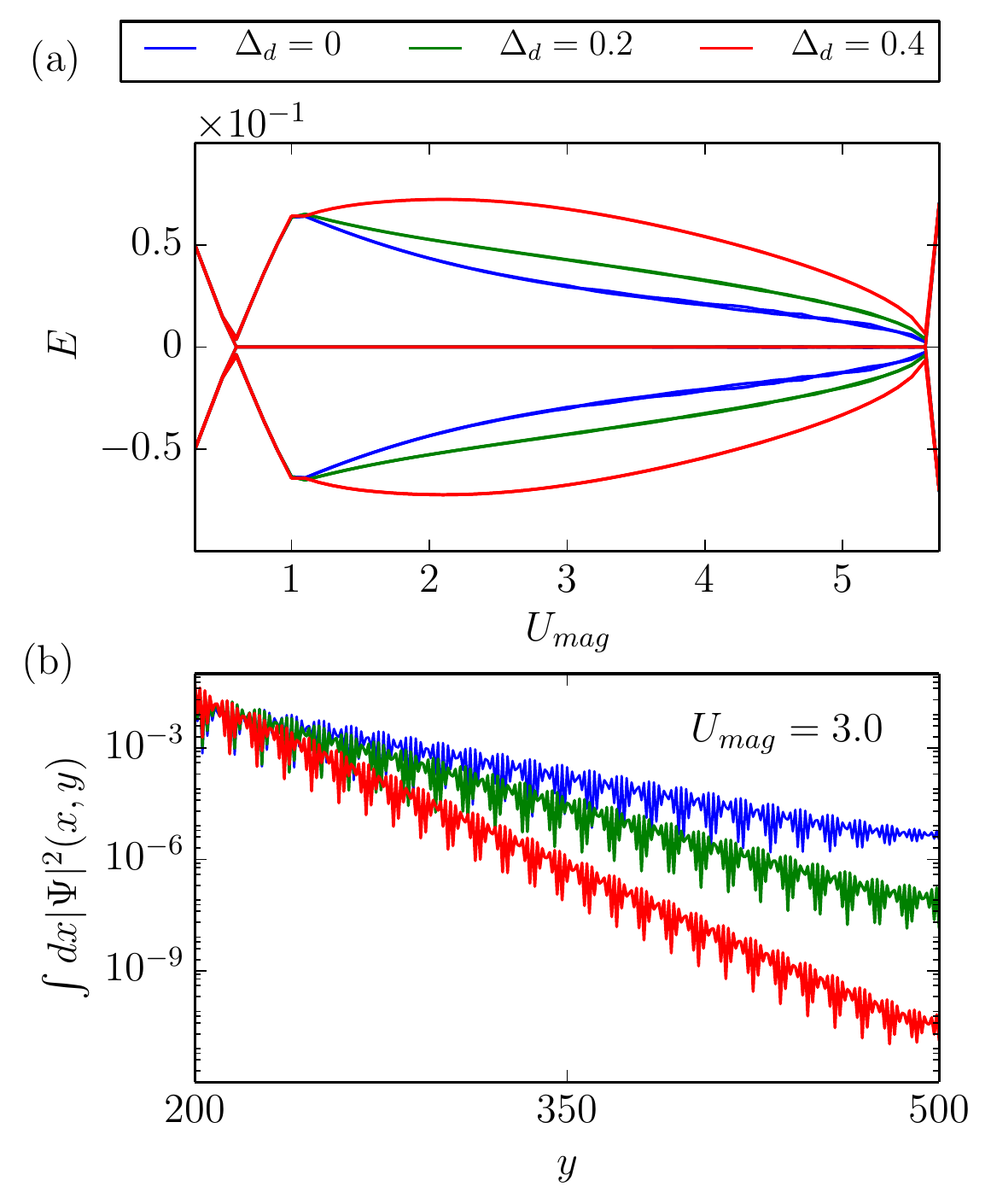}
\caption{(Color Online) Same as Fig.~\ref{d1+s_x.fig} but $d_{xy}{+}s$-wave SC for impurity chain oriented along $x$- or $y$-axis.}
\label{D_xy_pmS.fig}
\end{figure}

\subsection{$d_{xy}{+}s$-wave substrate}{\label{subsection_d2+s}}
Moving on to the other TRI $d$-wave substrate, $d_{xy}{+}s$, we use the same dimensional reduction scheme presented in Sec.~\ref{subsection_d1+s} and thus temporarily set $k_y=0$ $(k_x=0)$ for an impurity chain oriented along the $x$-axis ($y$-axis). For both chain orientations, the reduced order parameter reads $\Delta(k) = \Delta_s$. This means that for the $d_{xy}{+}s$-wave substrate, whether the chain is along the $x$- or $y$-axis, the order parameter is always finite. Another way to see this, is to look at the nodal lines of the order parameter in Fig.~\ref{NodalLine.fig}(b), where we can see that the $k_x=0$ and $k_y=0$ lines do not cross the nodal lines.   
Of course, the dimensional reduction analysis does not capture the changes in the order parameter due to the presence of $d$-wave order parameter and that is the main weakness of this analysis. However, this simple analysis has the benefit of predicting the invariance in chain orientation, which is remarkable. 

We present the energy of the lowest energy states for an impurity chain embedded in a $d_{xy}{+}s$-wave substrate in Fig.~\ref{D_xy_pmS.fig}(a), where we see the coexistence of $d_{xy}$-wave order with $s$-wave order leads to an increase in the minigap. As predicted by the dimensional reduction, we find the same  subgap states for $x$- and $y$-axis impurity chains. Moreover, as shown in Fig.~\ref{D_xy_pmS.fig}(b), we find an enhancement in the MBSs localization due to the presence of $d_{xy}$-wave order compared to pure $s$-wave SCs. We also notice that this coexistence does not change the critical coupling for the topological phase transition.
 
\begin{figure}[t]
\centering
\includegraphics[width=0.45\textwidth]{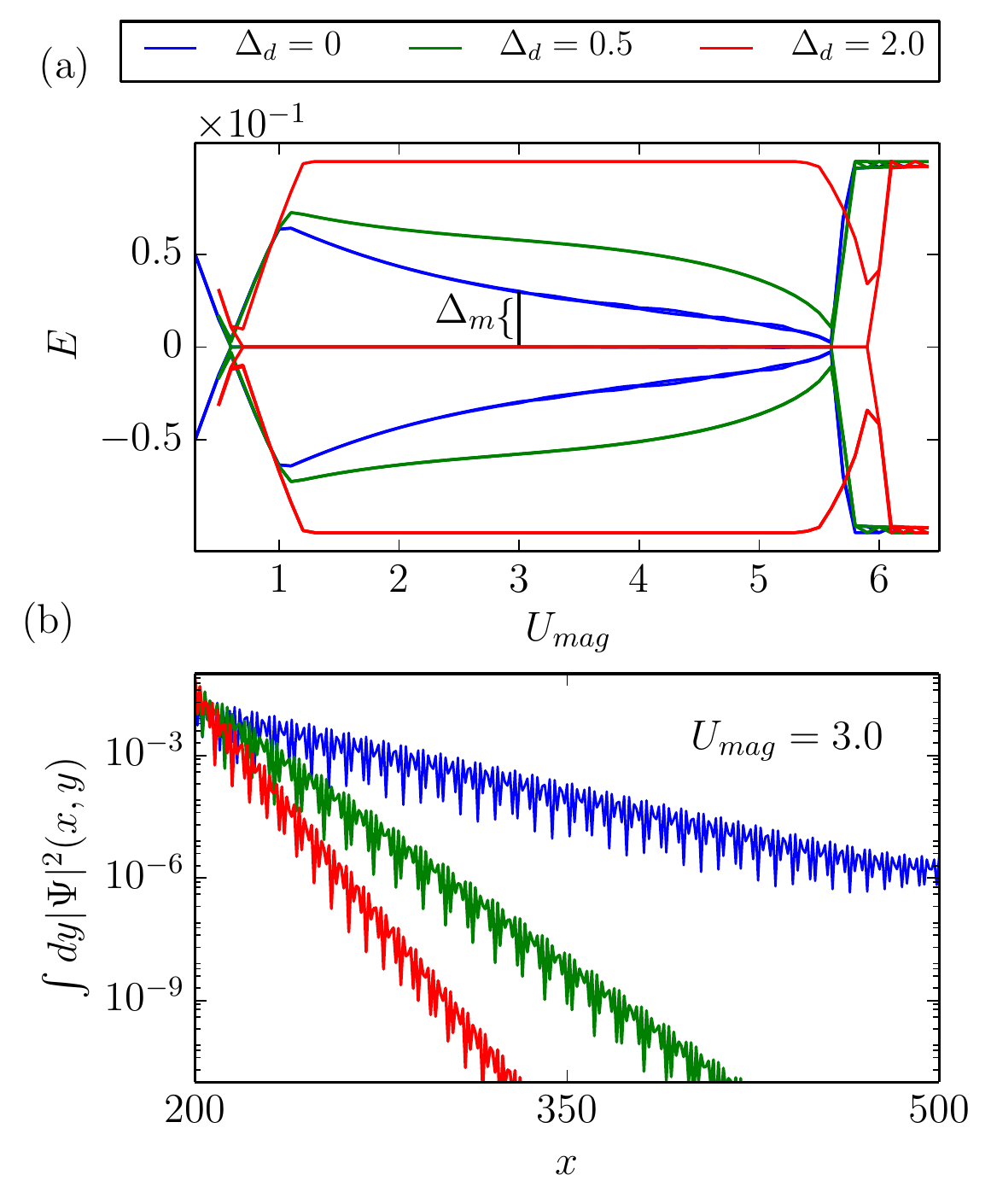}
\caption{(Color Online) Same as Fig.~\ref{d1+s_x.fig} but for $d_{x^2-y^2}{+}is$-wave SC impurity chain oriented along $x$- or $y$-axis.}
\label{D_x2-y2_pmiS.fig}
\end{figure}

\subsection{$d_{x^2-y^2}{+}is$-wave substrate}{\label{subsection_d1+is}}
We next turn to the TRB cases, where the order parameter develops a $\pi/2$ phase shift between the $s$- and $d$-wave parts. We start by studying the $d_{x^2-y^2}{+}is$-wave substrate. 
Just as before, we consider impurity chains orientated both along the $x$- or $y$-axis. 
As we calculate the energy of the lowest energy states for the TRB $d_{x^2-y^2}{+}is$-wave SC, we find the spectrum to be exactly the same for both chain orientations, see Fig.~\ref{D_x2-y2_pmiS.fig}(a). We can relate this orientation independence to the fact that the $d_{x^2-y^2}{+}is$-wave symmetry opens a hard gap in the spectrum due to the imaginary $s$-wave order parameter. Furthermore, the anisotropy does not make a difference between the $x$- and $y$-direction in terms of the magnitude of the gap, and thus $x$- and $y$-axis chains should both experience the same effective gap along the chain. 
The subgap state spectrum also reveals that adding the $d$-wave order parameter gives a larger minigap as well as much more localized MBSs, see Fig.~\ref{D_x2-y2_pmiS.fig}. Interestingly, in this TRB case, increasing $\Delta_d$ significantly increases $\Delta_m$ in the topological phase and therefore provides MBSs that survive at much higher temperatures. 
\begin{figure}[t]
\center
\includegraphics[width=0.45\textwidth]{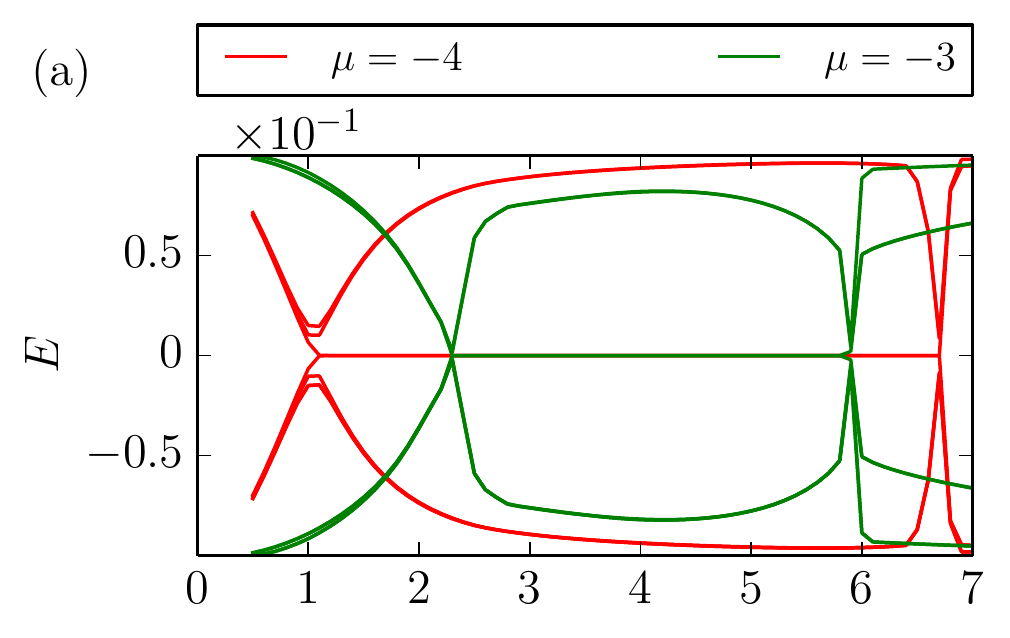}
\includegraphics[width=0.45\textwidth]{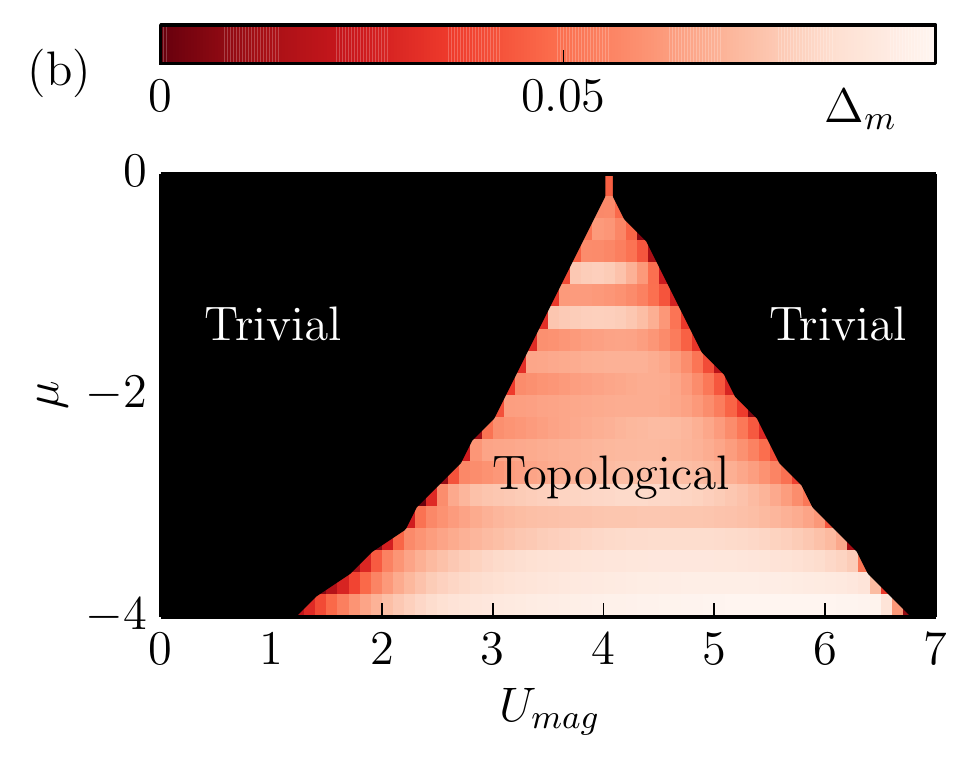}
\caption{(Color online) (a) Energy of lowest subgap states for $d_{x^2-y^2}{+}is$-wave SC with impurity chain oriented along $x$- or $y$-axis for $\mu=-3$ (green) and $\mu = -4$ (red). (b) Topological phase diagram for impurity chain in $d_{x^2-y^2}{+}is$-wave SC as a function of $\mu$ and $U_{\rm mag}$. Black region shows trivial phase; color scale represents minigap $\Delta_m$ in topologically nontrivial phase. Here $\Delta_d = 1$ and $\Delta_s=0.1$.}
\label{PhaseDiagram_d+is.fig}
\end{figure}

It is also important to notice that for the TRB $d_{x^2-y^2}{+}is$-wave state we do not have to fine-tune the chemical potential to bottom of the band $\mu=-4$ for the MBSs to emerge as the topological phase transition occurs for a wide range of chemical potentials. For example, Fig.~\ref{PhaseDiagram_d+is.fig}(a), where we plot the lowest energy states for $\mu=-3$ (green) and $\mu=-4$ (red), shows that the MBSs also appear for high doping where the minigap is also increased notably by an increasing $d$-wave component.
The physical origin of this tunability stems from the imaginary $s$-wave order parameter that, independent of any other normal state parameter, always opens a full energy gap. Therefore, the strong restriction on the $\Delta_d/\Delta_s$ ratio, chemical potential, and also Rashba spin-orbit coupling found for TRI substrate is lifted for TRB substrates. 

In Fig.~\ref{PhaseDiagram_d+is.fig}(b), we present the full topological phase diagram for an impurity chain in a $d_{x^2-y^2}{+}is$-wave SC, where the black regions represent the topologically trivial phase of the chain without any MBSs, while the triangular-shaped region shows the topologically nontrivial phase and its minigap. As is clearly seen, for all chemical potentials $\mu\in[-4,0]$ there exists a range of impurity strengths for which the system is in the topological phase. Due to the particle-hole symmetry of BdG Hamiltonian, the phase diagram for positive chemical potential $\mu\in[0,4]$ is given by simply flipping Fig.~\ref{PhaseDiagram_d+is.fig}(b) with respect to horizontal axis. Tuning the Rashba-spin orbit affects the minigap only slightly but does not change the shape of the phase diagram.

\begin{figure}[t]
\center
\includegraphics[width=0.4\textwidth]{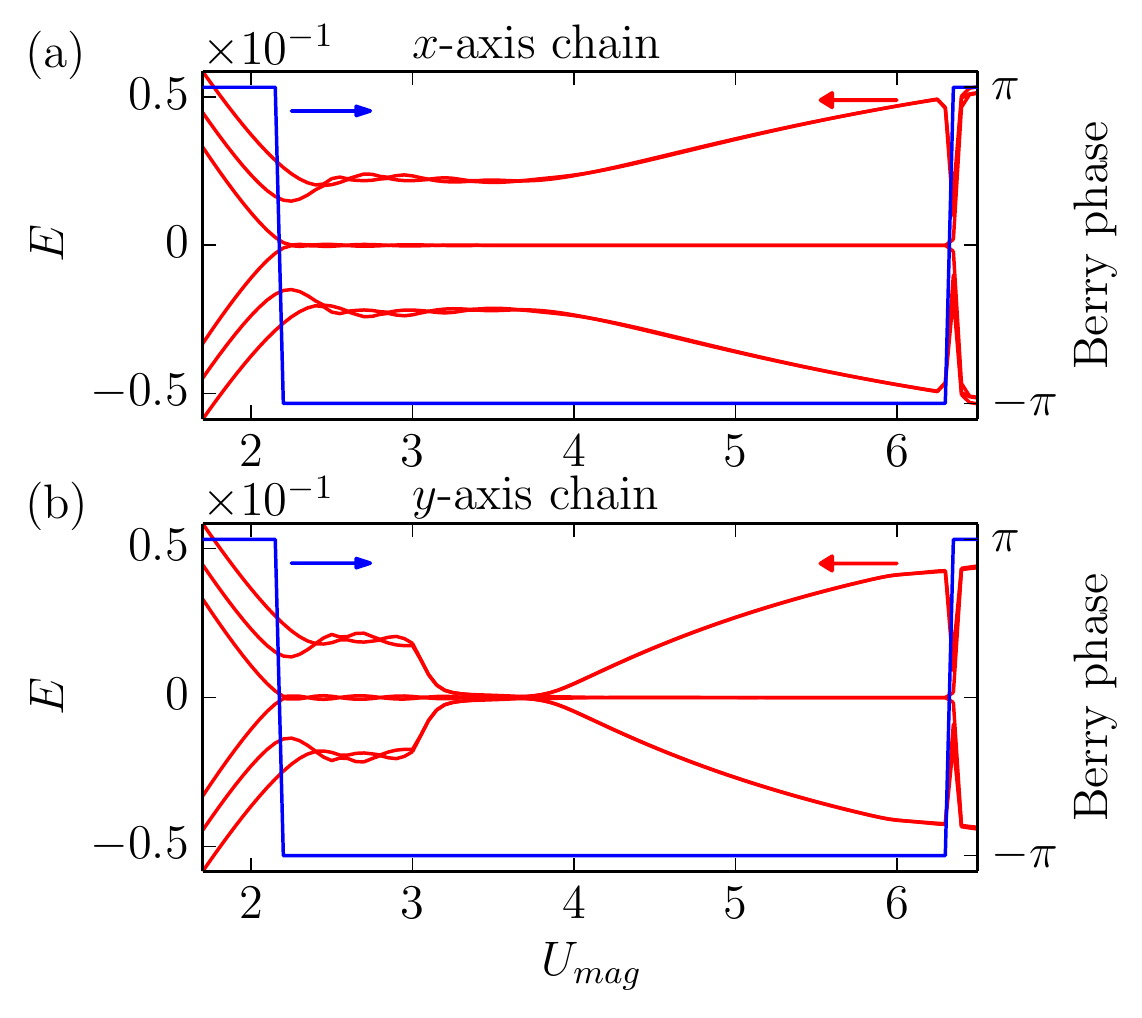}
\caption{(Color online) Spectrum of lowest subgap states (red, left axis) for $d_{xy}{+}is$-wave SC as a function of $U_{\rm mag}$ for impurity chains along $x$- (a) and $y$-axis (b) as well as Berry phase (blue, right axis). Here $t'= 0.1$, $\mu=-4$, and $\Delta_{d'}= 1$.}
\label{d2+is.fig}
\end{figure}

\subsection{$d_{xy}{+}is$-wave substrate}{\label{subsection_d2+is}}
We also consider the TRB order parameter of $d_{xy}{+}is$-wave symmetry with the impurity chain oriented along the $x$- or $y$-axis. Surprisingly, the orientation of the impurity chain in this case significantly affects the spectrum, in spite of the fact of having a hard gap and same magnitude of the order parameter along $x$- and $y$-directions. More precisely, the minigap for an impurity chain along the $x$-axis is different from a chain along the $y$-axis as depicted in Fig.~\ref{d2+is.fig}, where we plot the lowest energy states (red) for both chain directions. For the $x$-axis chain, we have also verified that the MBSs localization is enhanced due to the $d_{xy}$-wave order parameter in comparison to pure $s$-wave SC. 
However, when the impurity chain is along the $y$-axis, the energy spectrum exhibits more complexity. In this case, extra zero-energy states appear in an extra gap-closing in the middle of topological phase for intermediate coupling, $3 \lesssim U_{\rm mag} \lesssim 4$. We have verified that the chain end point MBSs exist independently of this extra gap-closing and when we introduce the next-nearest hopping, the coupling strength for which this extra gap-closing also changes, showing a model dependence. 

To assess the nature of these extra zero-energy states and extra gap-closing, we evaluate the Berry phase for the Fourier-transformed Hamiltonian along the chain using the Wilson loop formalism \cite{Bernevig2014, Adrein2018}. As the blue curve in Fig.~\ref{d2+is.fig} illustrates, we observe an abrupt change in the Berry phase between $\pi$ and $-\pi$ at $U_{\rm mag}^{(1)}$ and $U_{\rm mag}^{(2)}$. However, the Berry phase does not show any extra topological transition for $ 3\lesssim U_{\rm mag} \lesssim4$, which implies that the extra gap-closing in this region has a different origin rather than a topological phase transition.
\begin{figure}[t]
\center
\includegraphics[width=0.23\textwidth]{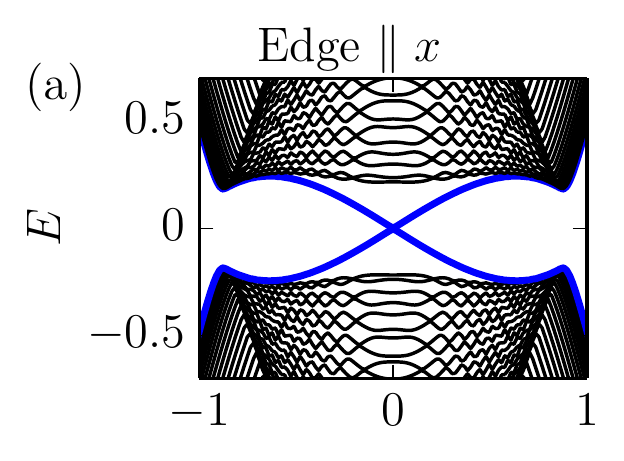}
\includegraphics[width=0.23\textwidth]{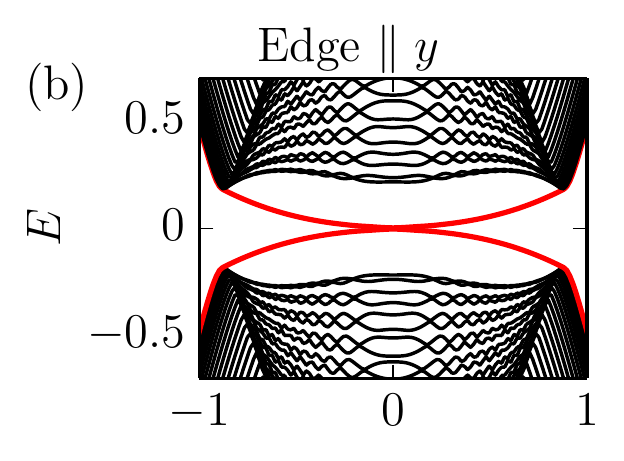}
\includegraphics[width=0.23\textwidth]{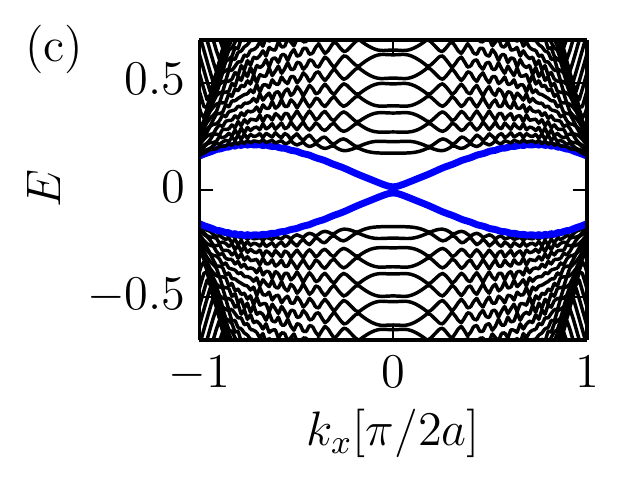}
\includegraphics[width=0.23\textwidth]{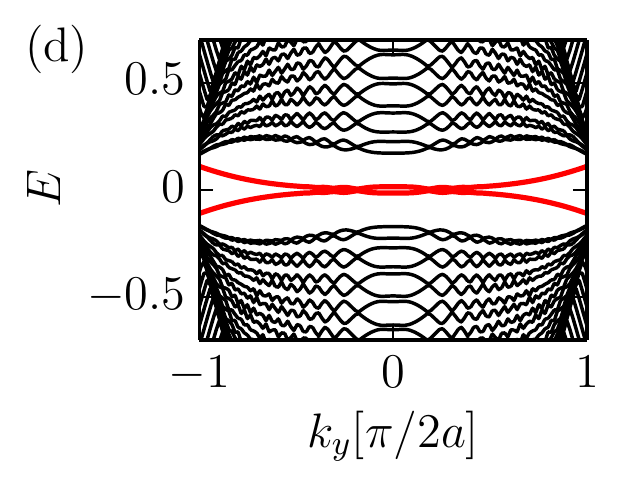}
\caption{(Color online) Chiral edge modes for nano-ribbon of $d_{xy}{+}is$-wave SC covered with a complete layer of magnetic impurities, with edge in $x$-direction (a), (c) and $y$-direction (b), (d). Here $\Delta_s=\Delta_{d'}= 0.5$ and  $U_{\rm mag} = 1.4$ (a), (b) and $U_{\rm mag} = 2.1$ (c), (d).}
\label{ML_D_xy_pm_iS.fig}
\end{figure}

In what follows, we perform a detailed analysis of the extra zero-energy states for the $y$-axis impurity chain but the lack thereof for an $x$-axis chain.
For this purpose, we first study a 2D spin-orbit coupled SC with $d_{xy}{+}is$-wave symmetry where the whole system is covered with magnetic impurities. The topological phase transition in similar systems has been studied previously for SCs with $s$-wave or $d{\pm}id'$-wave symmetries in the presence of external magnetic field \cite{Sato2010}. In principle, a system composed of a 2D SC covered with magnetic impurities can be seen as a parent model for the 1D impurity chain, since shrinking the magnetic cover only in one direction leads to the impurity chain embedded in a 2D SC. In the same fashion, the chiral edge states that appear at the edges of parent topological 2D SC become the MBSs that appear at the ends of the chain when shrinking the 2D impurity coverage to a 1D impurity chain. Notice how shrinking the impurity region in $x$- or $y$-directions gives an impurity chain along the $y$ or $x$-axis, respectively. 

To study the parent 2D model, we consider a superconducting nanoribbon (width 51 lattice points) fully covered with a layer of  magnetic impurities. We Fourier transform the Hamiltonian along the nanoribbon and observe that the system exhibits a topological phase transition with increasing $U_{\rm mag}$ into a topological phase with chiral edge modes, plotted in Fig.~\ref{ML_D_xy_pm_iS.fig}. 
Since the behavior of the low-energy states of the impurity chain depends on the chain orientation, we expect the chiral edge states in the parent 2D model to also show different dispersion relations on different edges. Remarkably, we find the Majorana chiral edge modes for the (10) edge (parallel to the $x$-axis) disperse differently from  edge modes on the (01) edge (parallel to the $y$-axis). Close to the $\Gamma$-point where band crossing takes place, the former has a linear, rather steep, dispersion relation, see Fig.~\ref{ML_D_xy_pm_iS.fig}(a) and \ref{ML_D_xy_pm_iS.fig}(c), while the latter displays a quadratic dispersion or even flatter as clearly seen in Fig.~\ref{ML_D_xy_pm_iS.fig}(b) and \ref{ML_D_xy_pm_iS.fig}(d). With increasing $U_{\rm mag}$, the edge states propagating along the $x$-axis have only one crossing at $\Gamma$ point while, for the modes propagating along the $y$-axis, several crossings appear and the edge states experience a very flat dispersion. 

To relate the 2D magnetic layer to the 1D impurity chain, we shrink the magnetic layer in one direction, which leads to discretization of the chiral edge states. One pair of these discrete energy levels sticks to zero energy, giving the MBSs, while the remaining nonzero energy levels are the YSR states. Therefore, when the chiral edge states along the $y$-axis become very flat, it means that, in addition to MBSs, there exist extra states in the middle of energy gap. These extra midgap states are not topologically protected but can still appear close to or even at zero energy. 
As a result, different dispersion relations for the 2D case gives very different low-energy spectra for $x$- and $y$-axis chains in the $d_{xy}{+}is$-wave SC, although both belong to the same topological class. This phenomena is connected to the way the chiral edge state's dispersion relation depends on the relation between the geometry of the boundary and the superconducting order parameter. Similar edge sensitivity has been seen in a chiral $p$-wave SC on the square lattice, where edge states disperse very different along the straight (10) and the zigzag (11) directions \cite{Bouhon14}. 
For the $d_{x^2-y^2}{+}is$-wave SC, there is no difference in 2D edge states, and thus $x$- and $y$-axis chains have the same low-energy spectrum.

\subsection{Self-consistent analysis}{\label{subsection_self-consistency}}
So far, we have assumed constant order parameter and neglected any depletion of the order parameter in the vicinity of the impurity chain and its consequences. To assess impurity chains while relaxing this constraint, we also perform reference self-consistent calculations for the superconducting order parameter. Here, we only have to assume a finite and constant pair potential $V$ in each pairing channel but then calculate the order parameter(s) explicitly everywhere in the lattice. For a $d$-wave state, we use the self-consistent condition  $\Delta_d {(\bf{i},\bf{j})} = -V_d/2\langle  c_{\bf{i}\downarrow}c_{\bf{j}\uparrow} -   c_{\bf{i}\uparrow}c_{\bf{j}\downarrow} \rangle$, where $\bf{i},\bf{j}$ are nearest-neighbor sites. In the self-consistent calculation, we start by guessing a value for $\Delta_d$ on each bond, solve Eq.~\eqref{BdG1}, evaluate a new $\Delta_d$ on each bond using the self-consistent condition, and repeat until $\Delta_d$ does not change between two subsequent iterations. We emphasize that the order parameter on vertical and horizontal bonds is solved independently to also allow for the system to chose the competing extended $s$-wave symmetry.  We also assume a finite $V_s$ in addition to $V_d$ and separately calculate  $\Delta_s = -V_s/2\la c_{\bf{i}\downarrow} c_{\bf{i}\uparrow} - c_{\bf{i}\uparrow} c_{\bf{i}\downarrow} \ra$ self-consistently. The phase difference between $\Delta_d$ and $\Delta_s$ is also found self-consistently, i.e.,~we only start with a specific phase difference, but then let the system evolve without any constraints. 

As an example, we take an impurity chain along the $x$-axis on the surface of $d_{x^2-y^2}{+}is$-wave SC and find all the order parameters self-consistently. For $\mu=-2$, we find the $\pi/2$ phase-shift between the $s$-wave and $d_{x^2-y^2}$-wave order parameters even in the fully self-consistent solution.
We see that, in this case, the dominant $d_{x^2-y^2}$-wave and subdominant $s$-wave order parameters are both heavily depleted in the vicinity of the impurity chain and a small extended $s$-wave order parameter also appears close to the chain, similar to the situation for a single magnetic impurity \cite{Mahdi2017}. 
Still, in the topological phase the minigap and the localization of the MBSs is enhanced by $d_{x^2-y^2}$-wave order parameter, very similarly to the non-self-consistent results reported earlier in Sec.~\ref{subsection_d1+is}.
Self-consistency does move the critical coupling for which the topological phase transition takes place to lower values, but the size of the topological region, namely the region between gap-closing and gap-reopening, is not affected by self-consistency. Therefore, we conclude that self-consistency does not change the conclusions drawn earlier with non-self-consistent calculations. 

\begin{figure}[b]
\center
\includegraphics[width=0.5\textwidth]{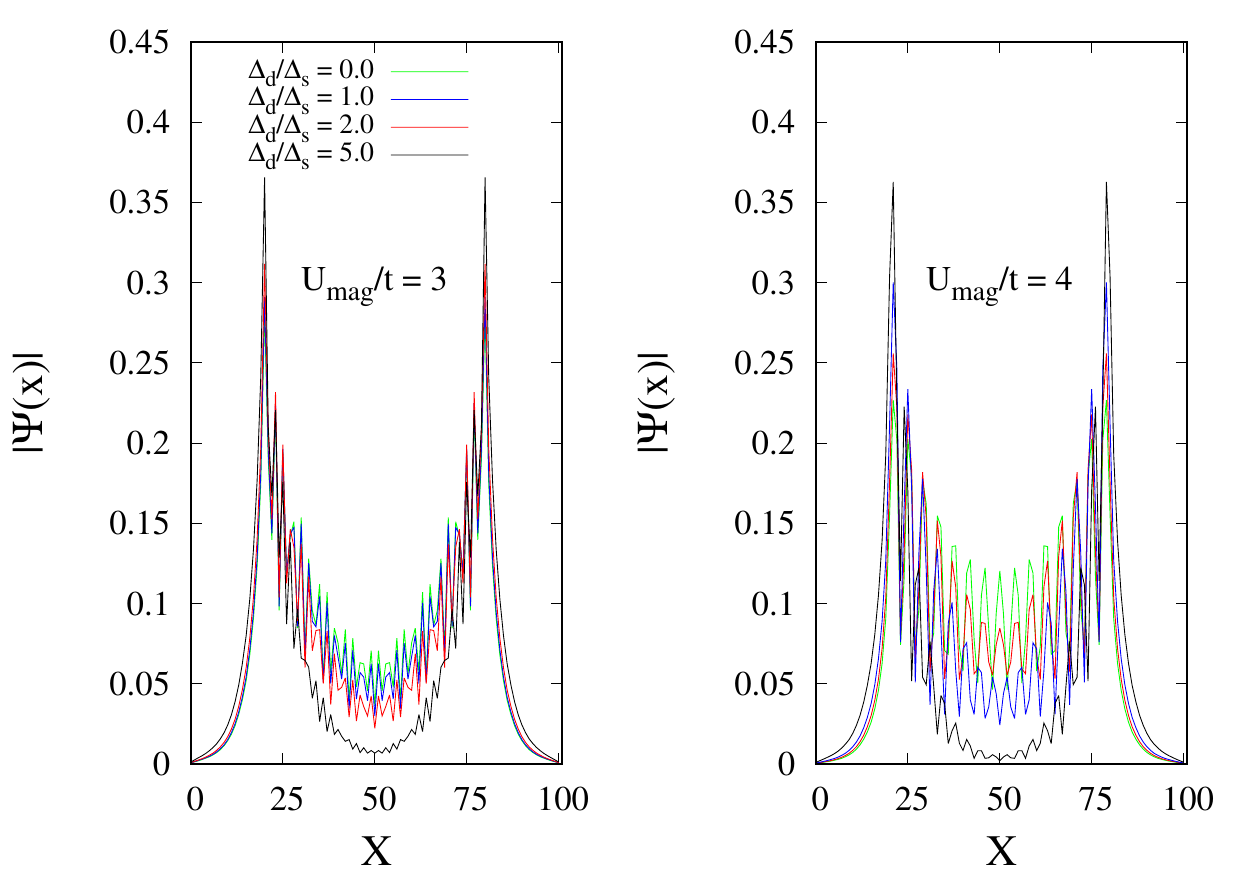}
\caption{(Color Online) Absolute value of MBS's wave function for a spin helix structure without any Rashba spin-orbit coupling in substrate. Spin helix lay in the $x-y$ plane and spins rotates with a pitch of $k_ha=2\pi/3$ along the chain.}
\label{SpinHelix.fig}
\end{figure}
\subsection{Spin helical impurity chain}{\label{subsection_spin-helix}}
To assess the generality of the obtained results using ferromagnetic chains, we also study a spin helical impurity chain. Here we exclude Rashba spin-orbit interaction in the substrate and instead assume an in-plane spin-helix structure for the local moments of the impurities \cite{Choy2011,Klinovaja2013,Nadj-Perge2013,Oppen2014}. We choose a pitch of $k_ha=2\pi/3$ along an $x$-axis impurity chain and no out-of-plane spin component, but the results are not sensitive to this particular choice. In Fig.~\ref{SpinHelix.fig}, we plot the absolute value of the MBS's wave function, assuming a $d_{x^2-y^2}{+}is$-wave substrate. We notice that increasing the ratio of $\Delta_d/\Delta_s$ leads to more localized MBSs, similar to the effect for the ferromagnetic impurity chain. The outcome of this calculation, thus, reveals that our results are also generally applicable to spin-helix structures.

%
%
\section{Discussion}{\label{section_discussion}}
Having, in detail, analyzed the different combinations of $d$- and $s$-wave orders in the preceding sections and especially how a $d$-wave state can enhance the robustness of the MBSs, we summarize the results in Fig.~\ref{mini-gap.fig}. Here we plot the minigap for all studied condensates and chain directions as a function of the ratio $\Delta_d/\Delta_s$. For $x$-axis chains and TRI SC, we see in Fig.~\ref{mini-gap.fig}(a) that the minigap is enhanced by increasing the $d$-wave order all the way to $\Delta_d/\Delta_s \lesssim 5$. However, for very large values of $\Delta_d/\Delta_s$ the minigap is suppressed and eventually vanishes due to nodes in the energy spectrum then appearing in the vicinity of the chain. In contrast, as shown in Fig.~\ref{mini-gap.fig}(b), for any TRB substrate the minigap is enhanced monotonously with an increasing $d$-wave component, eventually saturating at $\Delta_m=\Delta_s$, i.e.,~much larger than the minigap in the pure $s$-wave case. Consequently, a $d$-wave order parameter enhances the minigap and thus the robustness of the MBSs for $x$-axis chains over a wide range of $\Delta_d/\Delta_s$ ratios for all types of condensates.

Turning to $y$-axis chains embedded in TRI SC as shown in Fig.~\ref{mini-gap.fig}(c), we see that the minigap for $d_{xy}{+}s$-wave SC is orientation independent and exactly similar to $x$-axis chain in Fig.~\ref{mini-gap.fig}(a). On the other hand, for the TRI $d_{x^2-y^2}{+}s$-wave SC, the $d$-wave order does not enhance the minigap since the chain orientation crosses the nodal lines of order parameter. For $y$-axis impurity chains in a TRB substrate, we see in Fig.~\ref{mini-gap.fig}(d) that when the substrate has $d_{x^2-y^2}+is$-wave symmetry, the minigap behaves exactly similar to the $x$-oriented chain in Fig.~\ref{mini-gap.fig}(c). When the substrate is a $d_{xy}{+}is$-wave SC, we, in addition, find the exotic minigap closing explained in Sec.~\ref{subsection_d2+is}, which for a small range of $\Delta_d/\Delta_s$ ratios suppresses the minigap that is otherwise notably enhanced over the pure $s$-wave case.
\begin{figure}[t]
\center
\includegraphics[width=0.5\textwidth]{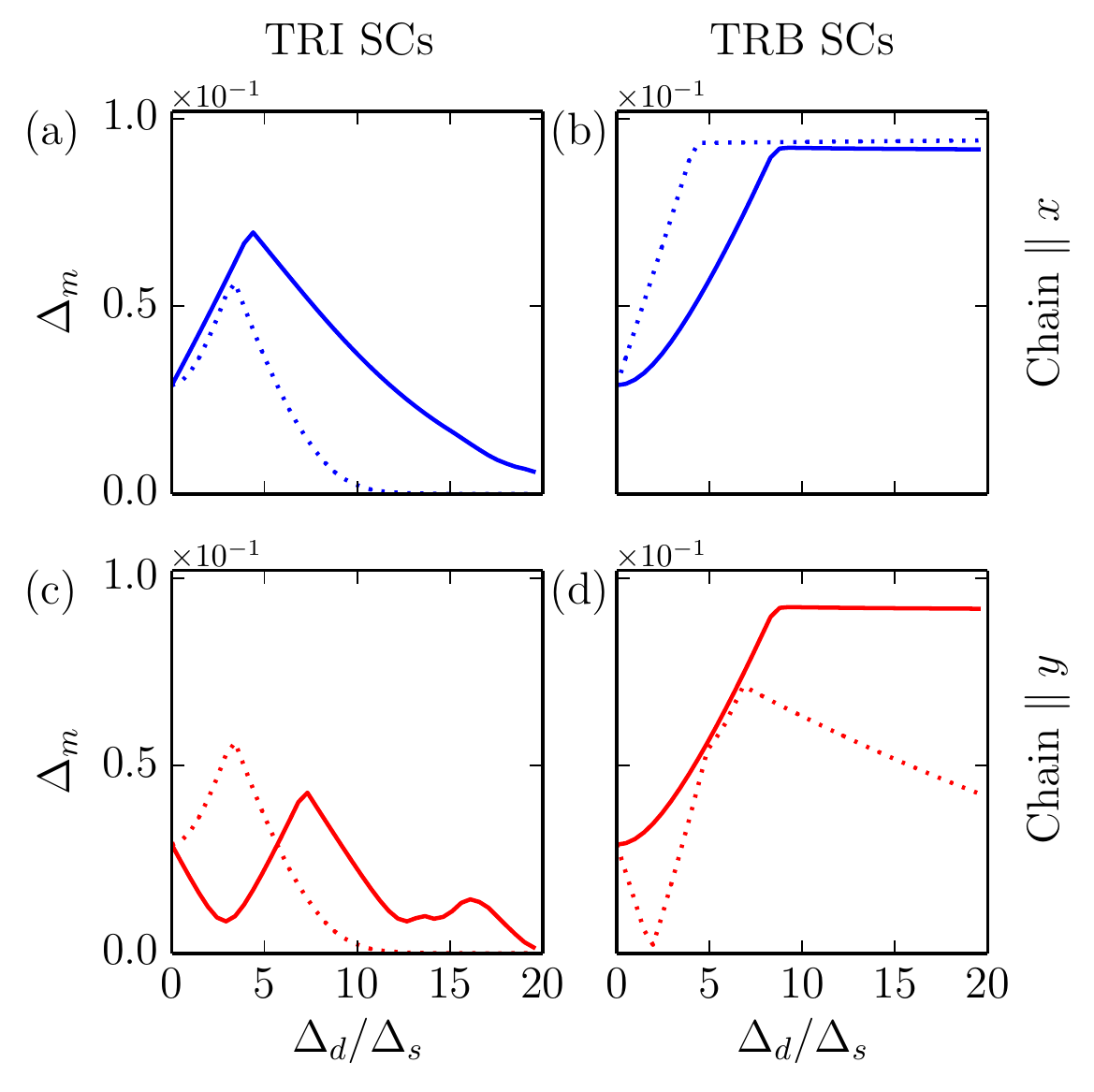}
\caption{(Color Online) Minigap $\Delta_m$ for $x$- (a), (b) and $y$-axis (c), (d) impurity chains embedded in different superconducting substrates as a function of $\Delta_d/\Delta_s$ ratio. For TRI (TRB) SCs when the $d$-wave order has $d_{x^2-y^2}{+}s$-wave ($d_{x^2-y^2}{+}is$-wave) symmetry, $\Delta_m$ is solid blue (red), while for $d_{xy}{+}s$-wave ($d_{xy}{+}is$-wave) symmetry $\Delta_m$ is dotted blue (red). Here $U_{\rm mag}=3$, $\mu=-4$, $\Delta_s = 0.1$.}
\label{mini-gap.fig}
\end{figure}
Considering the topological classification of the superconducting states, impurity chains embedded in a pure $s$-wave SC and in combinations of $d$- and $s$-wave SCs belong to the same topological class and thus the emerged MBSs possess the same overall properties and non-Abelian statistics. Here, spin-polarized scanning tunneling spectroscopy can provide a suitable experimental tool to detect the MBSs  and differentiate it from other (near) zero-energy states \cite{Jeon17}.

The results summarized in Fig.~\ref{mini-gap.fig} show how a $d$-wave order parameter is often highly beneficial for MBS robustness. Still, if the $d$-wave order becomes extremely dominant and has nodes crossing close to a poorly chosen chain direction, the minigap is  reduced and the MBS eventually disappears.
One might argue that the virtually bound resonance states in pure $d$-wave SCs, which can appear at zero energy \cite{DM}, are spin polarized, and can thus substitute the YSR states that ultimately produce the MBSs in the topological phase. 
Although zero-energy end states theoretically appear in even the pure $d$-wave case, our calculations reveals that in the absence of an $s$-wave order and even for very large $d$-wave order parameter, the minigap is extremely small $\Delta_{m}/\Delta_{d_{x^2-y^2}} \le 10^{-3}$ and $\Delta_{m}/\Delta_{d_{xy}} \le 10^{-8}$. Thus, even if these zero-energy modes are technically zero-energy states, they are empirically hybridized even at very low temperatures and can hardly be utilized as MBSs.

Considering the experimental realization of an impurity chain platform in $d$-wave SCs, cuprate surfaces with its proposed TRB phase is an interesting alternative. For example, the $d+is$-wave state has been identified in nanoislands of YBa$_2$Cu$_3$O$_{7-\delta}$. Alternatively, a hybrid structure of unconventional $d$-wave SC and a thin layer of a conventional $s$-wave SC can produce a superconducting state where $d$-wave state coexists with a required additional $s$-wave component. 
In this case, the impurity chain can, e.g.,~be located on the surface of the $d$-wave SC and the tunneling probe measurement can be done on the outer $s$-wave layer. A version of such a setup has been recently used to study the impurity Co islands under a single layer of Pb SC \cite{Menard2017}.
Note also that the reflection symmetry is broken at surfaces/interfaces, which then automatically provide the Rashba spin-orbit coupling which is essential for the  MBSs to emerge.

Finally, let us compare our results with the case of a semiconductor nanowire in proximity to a spin-orbit-coupled $d$-wave SC.
In Refs.~\cite{Takei2013,Ortiz2018}, superconductivity is proximity induced into the nanowire and with the electronic bands in the wire being spin-polarized, the pairing is actually in the spin-triplet channel. The model employed in this paper and the one used in Refs.~\cite{Takei2013,Ortiz2018} are thus aimed to explain different experimental set-ups and the results for these two models are thus not always similar. For instance, both works predict that the properties of MBSs can be direction dependent \cite{Takei2013}. However, in modeling the nanowire, the localization of MBSs was shown to be reduced due to the angular asymmetry of the $d$-wave order in large regions of topological phase when compared to a nanowire on top of a conventional $s$-wave SC using $\Delta_d = \Delta_s$ \cite{Ortiz2018}. In contrast, for the emergence of MBSs in an impurity chain in a $d$-wave SC, we show that we need coexistence of $s$-wave and $d$-wave order parameters and actually observe much more localized MBSs for $d_{x^2-y^2}{+}is$-wave SC than in the conventional $s$-wave case.

%
%

\section{Conclusions}{\label{section_conclusion}}
In this paper, we study a chain of magnetic impurities located on the surface of a $d$-wave SC with a subdominant $s$-wave order parameter and in the presence of Rashba spin-orbit coupling from inversion-breaking surface. This set-up is a promising platform for realizing MBSs and exploiting their non-Abelian statistics in high-temperature SCs. Performing numerical tight-binding lattice calculations, we investigate the effect of $d$-wave pairing on the topological phase transition and the associated MBSs. We show that a pair of MBSs emerge at the two end points of the impurity chain for a wide range of physical parameters and for both TRI and TRB condensates. The presence of the $d$-wave order parameter provides the advantage of larger order parameter thanks to higher superconducting transition temperature. Remarkably, as long as the chain orientation does not cross any remaining nodal lines of the order parameter, the presence of the $d$-wave order gives rise to dramatically more localized MBSs than the pure $s$-wave case. This we attribute to the large enhancement of the effective order parameter along the impurity chain.
We also show that the $d$-wave order parameter can strongly enhance the minigap energy which protects the MBSs from thermal hybridization. Larger minigap offers a promising way to increase the robustness of MBSs specially for a TRB substrate. This property should not be confused with the localization of MBSs since we bring an example where more localized MBSs emerge with smaller minigap.  

Furthermore, we report on an exotic feature for an impurity chain along the $y$-axis and embedded in a $d_{xy}{+}is$-wave SC, where an extra gap-closing occurs within the topologically nontrivial phase. Evaluating the Berry phase, we do not find any signature for a topological phase transition at this extra gap-closing point. Instead, we trace the extra gap-closing back to a flat dispersion of the topological edge state of the equivalent 2D system. This result shows that even 1D topological phases can exhibit a low-energy spectrum not determined by topology alone.
To conclude, this paper shows that using a $d$-wave SC with any subdominant $s$-wave order can strongly enhance the thermal robustness and localization of MBSs. This paves the way for topological quantum computation at much higher temperatures and will hopefully inspire both future experimental and theoretical investigation in this direction.

\acknowledgments
We thank K.~Bj{\"o}rnson and A.~Bouhon for fruitful discussions and J.~Cayao and F.~Parhizgar for valuable comments on the paper.
We acknowledge financial support from the Swedish Research Council (Vetenskapsr\aa det, Grant No.~621-2014-3721), the G\"{o}ran Gustafsson Foundation, the Swedish Foundation for Strategic Research (SSF), and the Wallenberg Academy Fellows program and the Knut and Alice Wallenberg Foundation.

\bibliography{bibFile.bib}

\begin{thebibliography}{65}
\bibitem [{Myn()}]{Mynote1}%
  \BibitemOpen
  \href@noop {} {\bibinfo  {journal} {An accurate dimensional reduction should
  integrate over the direction perpendicular to the chain, which is beyond the
  scope of this paper}\ }\BibitemShut {NoStop}%
\end{thebibliography}

\end{document}